\documentclass{article}
\usepackage{times}
\usepackage{amsfonts}
\usepackage{graphicx}
\usepackage[pdfmark]{hyperref}
\begin{document}
\noindent
{\Large THE RICCI FLOW ON RIEMANN SURFACES}
\vskip1cm
\noindent
{\bf S. Abraham}${}^{1}$, {\bf P. Fern\'andez de C\'ordoba}${}^{2}$, {\bf Jos\'e M. Isidro}${}^{3}$ and {\bf J.L.G. Santander}${}^{4}$\\
Grupo de Modelizaci\'on Interdisciplinar, Instituto de Matem\'atica Pura y Aplicada,\\ Universidad Polit\'ecnica de Valencia, Valencia 46022, Spain\\
${}^1${\tt sabraham@mat.upv.es}, ${}^2${\tt pfernandez@mat.upv.es}, \\
${}^3${\tt joissan@mat.upv.es}, ${}^4${\tt jlgonzalez@mat.upv.es}
\vskip1cm
\noindent
{\bf Abstract} We establish a 1--to--1 relation between metrics on compact Riemann surfaces without boundary, and mechanical systems having those surfaces as configuration spaces.

\section{Introduction}\label{intt}

\subsection{Motivation}\label{motius}

Let $\mathbb{M}$ be a smooth $n$--dimensional manifold endowed with the local coordinates $q_i$, $i=1,\ldots, n$, that we regard as the configuration space of some classical mechanical system with the Lagrangian function $L$,
\begin{equation}
L=T-V=\frac{1}{2}a_{ij}(q)\dot q_i\dot q_j-V(q).
\label{fermat}
\end{equation}
Here $V$ denotes the potential energy of the system, and $T$ is the kinetic energy (a positive definite quadratic form in the velocities $\dot q_i$). Using these data we can construct a Riemannian metric as follows.  Consider the momenta $p_i$ conjugate to the $q_i$,
\begin{equation}
p_i(q)=\frac{\partial L}{\partial \dot q_i}=a_{ij}(q)\dot q_j.
\label{memo}
\end{equation}
Then the 1--form
\begin{equation}
p_i{\rm d}q_i=a_{ij}\dot q_j{\rm d}q_i=\frac{1}{{\rm d}t}a_{ij}{\rm d}q_i{\rm d}q_j,
\label{foolish}
\end{equation}
is the integrand of Hamilton's principal function (or time--independent action):
\begin{equation}
S[q]:=\int p_i{\rm d}q_i.
\label{prin}
\end{equation}
Now conservation of energy implies that the Hamiltonian function $H$,
\begin{equation}
H=\frac{1}{2}a_{ij}\dot q_i\dot q_j+V(q),
\label{sinatra}
\end{equation}
is a constant of the motion, that we denote by $E$. We can solve (\ref{sinatra}) for the square root of the quadratic form,
\begin{equation}
\sqrt{a_{ij}{\rm d}q_i{\rm d}q_j}=\sqrt{2(E-V(q))}\,{\rm d}t,
\label{savings}
\end{equation}
and substitute the result into (\ref{prin}) after using (\ref{foolish}), to find
\begin{equation}
S[q]=\int \sqrt{2(E-V(q))}\sqrt{a_{ij}{\rm d}q_i{\rm d}q_j}=:\int{\rm d}s.
\label{aktion}
\end{equation}
Determining the actual trajectory followed by the particle is therefore equivalent to finding the shortest path between two given points, with distances measured with respect to the (square root of the) quadratic form d$s^2$:
\begin{equation}
{\rm d}s^2:=g_{ij}(q){\rm d}q_i{\rm d}q_j, \qquad g_{ij}(q):=2(E-V(q))a_{ij}(q)
\label{metrix}
\end{equation}
The factor $2(E-V(q))$ is positive away from those points at which the particle is at rest (where $T=0$,  hence $E=V(q)$). Let $\mathbb{M}'$ denote the subset of all points of $\mathbb{M}$ at which the particle is {\it not}\/ at rest:
\begin{equation}
\mathbb{M}':=\left\{q\in \mathbb{M}: T\vert_q>0\right\}.
\label{sotto}
\end{equation}
We will assume that $\mathbb{M}'$ qualifies as a manifold (possibly as a submanifold of $\mathbb{M}$), and that the matrix $a_{ij}(q)$ is everywhere nondegenerate on $\mathbb{M}'$. This implies that $g_{ij}(q)$ is nondegenerate on $\mathbb{M}'$. Moreover, the quadratic form $a_{ij}(q)$ is positive definite and symmetric. Altogether, $\mathbb{M}'$ qualifies as a Riemannian manifold. On the latter, determining the actual trajectories for the particle is equivalent to determining the geodesics of the metric (\ref{metrix}).

Of course, all of the above is well known in the literature \cite{ARNOLD}.  The statement that the actual motion of the particle follows the geodesics of the metric (\ref{metrix}) goes by the name of {\it Fermat's principle}\/ (see, {\it e.g.}, ref. \cite{BUEHLER} for a nice account). The latter is equivalent to the principle of least action in Lagrangian mechanics. In this paper we address the converse problem, namely: to determine a point mechanics starting from the knowledge of a Riemannian metric on a given manifold. The sought--after mechanical system must somehow be canonically associated with the given metric, in the sense that it must be a {\it natural choice}\/, so to speak, among all possible point mechanics that one can possibly define on the given Riemannian manifold.

This problem will turn out to be too hard to solve in all generality---if it possesses a solution at all. Indeed, on an $n$--dimensional manifold, a general metric is determined (in local coordinates) by the knowledge of $n(n+1)/2$ coefficient functions $g_{ij}$, out of which some potential function $U$ and some positive--definite kinetic energy $T$ must be concocted. We can, however, make some simplifying assumptions. An educated guess leads us to restrict our attention to 2--dimensional manifolds $\mathbb{M}$, the simplest on which nontrivial metrics can exist. On the latter class of manifolds, any Riemannian metric is conformal, so it is univocally determined by the knowledge of just one function, the so--called {\it conformal factor}\/. Having got this far we can unashamedly declare $\mathbb{M}$ (our would--be configuration space) to be a compact Riemann surface without boundary. Compactness ensures the convergence of the integrals we will work with, without the need to impose further conditions on the integrands  (such as, {\it e.g.}, fast decay at infinity). The absence of a boundary ensures the possibility of integrating by parts without picking up boundary terms. However it should be realised that imposing these two requirements (compactness and the absence of a boundary) is a useful, but by no means necessary, condition to achieve our goal, namely: to relate metrics on configuration space with mechanical models on that same space.  Contrary to the previous example, we will not require that geodesics of the metric be actual trajectories for the mechanics. This notwithstanding, interesting links between mechanics and geometry will be exposed.

We should point out that there is, of course, a natural choice of a mechanics for a given family of metrics on a manifold---namely the one defined by the Einstein--Hilbert gravitational action functional. However the latter defines not a point mechanics, but a field theory. Moreover, this field theory has the space of all metrics on $\mathbb{M}$ as its configuration space. We are interested in a point mechanics, the configuration space of which is the Riemann surface  $\mathbb{M}$. Surprisingly, the point mechanics we will construct will be intimately related with the gravitational action functional.

\subsection{Setup}\label{gauss}

On our compact Riemann surface without boundary $\mathbb{M}$ there exist isothermal coordinates $x,y$, in which the metric reads \cite{FARKASKRA}
\begin{equation}
g_{ij}={\rm e}^{-f}\delta_{ij}, \qquad {\rm d}s^2={\rm e}^{-f(x,y)}({\rm d}x^2+{\rm d}y^2)
\label{metrik}
\end{equation}
where $f=f(x,y)$ is a function, hereafter referred to as {\it conformal factor}\/. The volume element on $\mathbb{M}$ equals\footnote{Our conventions are $g=\vert\det g_{ij}\vert$ and $R_{im}=g^{-1/2}\partial_n\left(\Gamma_{im}^ng^{1/2}\right)-\partial_i\partial_m\left(\ln g^{1/2}\right)-\Gamma_{is}^r\Gamma_{mr}^s$ for the Ricci tensor, $\Gamma_{ij}^m=g^{mh}\left(\partial_ig_{jh}+\partial_jg_{hi}-\partial_hg_{ij}\right)/2$ being the Christoffel symbols.}
\begin{equation}
\sqrt{g}\,{\rm d}x{\rm d}y={\rm e}^{-f}{\rm d}x{\rm d}y. 
\label{skalar}
\end{equation}
Given an arbitrary function $\varphi(x,y)$ on $\mathbb{M}$, we have the following expressions for the Laplacian $\nabla^2\varphi$ and the squared gradient $\left(\nabla\varphi\right)^2$:
\begin{equation}
\nabla^2\varphi:=\frac{1}{\sqrt{g}}\partial_m\left(\sqrt{g}g^{mn}\partial_n\varphi\right)={\rm e}^f\left(\partial_x^2\varphi+\partial_y^2\varphi\right)=:{\rm e}^fD^2\varphi,
\label{eins}
\end{equation}
\begin{equation}
\left(\nabla\varphi\right)^2:= g^{mn}\partial_m\varphi\partial_n\varphi={\rm e}^f\left[\left(\partial_x\varphi\right)^2+\left(\partial_y\varphi\right)^2
\right]=:{\rm e}^f\left(D\varphi\right)^2,
\label{drei}
\end{equation}
where $D^2\varphi$ and $\left(D\varphi\right)^2$ stand for the flat--space values of the Laplacian and the squared gradient, respectively. The Ricci tensor reads 
\begin{equation}
R_{ij}=\frac{1}{2}D^2f\,\delta_{ij}=\frac{1}{2}{\rm e}^{-f}\nabla^2f\,\delta_{ij}.
\label{ritchie}
\end{equation}
{}From here we obtain the Ricci scalar
\begin{equation}
R={\rm e}^{f}D^2f=\nabla^2f.
\label{rico}
\end{equation}
Now Perelman's functional ${\cal F}[\varphi,g_{ij}]$ on the Riemann surface $\mathbb{M}$ is defined as \cite{PERELMAN, TOPPING}
\begin{equation}
{\cal F}[\varphi,g_{ij}]:=\int_{\mathbb{M}}{\rm e}^{-\varphi}\left[\left(\nabla \varphi\right)^2+ R(g_{ij})\right]\sqrt{g}\,{\rm d}x{\rm d}y.
\label{epe}
\end{equation}
By (\ref{rico}) we can express ${\cal F}[\varphi,g_{ij}]$ as 
\begin{equation}
{\cal F}[\varphi,f]:={\cal F}[\varphi,g_{ij}(f)]=\int_{\mathbb{M}}{\rm e}^{-\varphi-f}\left[\left(\nabla \varphi\right)^2+\nabla^2f\right]{\rm d}x{\rm d}y.
\label{epep}
\end{equation}
The gradient flow of ${\cal F}$ is determined by the evolution equations
\begin{equation}
\frac{\partial g_{ij}}{\partial t}=-2\left(R_{ij}+\nabla_i\nabla_j\varphi\right),\qquad \frac{\partial\varphi}{\partial t}=-\nabla^2\varphi-R.
\label{rancio}
\end{equation}
Via a time--dependent diffeomorphism, the above are equivalent to 
\begin{equation}
\frac{\partial g_{ij}}{\partial t}=-2R_{ij}, \qquad \frac{\partial \varphi}{\partial t}=-\nabla^2\varphi+\left(\nabla\varphi\right)^2-R.
\label{muyrancio}
\end{equation}
Setting now $\varphi=f$ in (\ref{epep}) we have
\begin{equation}
{\cal F}[f]:={\cal F}[\varphi=f,f]=\int_{\mathbb{M}}{\rm e}^{-2f}\left[(\nabla f)^2+\nabla^2f\right]{\rm d}x{\rm d}y,
\label{todorancio}
\end{equation}
and the second eqn. in (\ref{muyrancio}) becomes, by (\ref{rico}),
\begin{equation}
\frac{\partial \tilde f}{\partial t}+2\nabla^2 \tilde f-\left(\nabla \tilde f\right)^2=0.
\label{konvfluss}
\end{equation}
In the time--flow eqn. (\ref{konvfluss}) we have placed a tilde on top of the conformal factor in order to distinguish it from the time--independent $f$ present in the functional (\ref{todorancio}). This improvement in notation will turn out to be convenient later on.

\subsection{Summary of results}\label{satz}

{\bf Theorem.} {\it Let $\mathbb{M}$ be a compact Riemann surface without boundary, and regard $\mathbb{M}$ as the configuration space of a classical mechanical system, with a potential $U$ proportional to the Ricci scalar curvature of $\mathbb{M}$. Then there exists a 1--to--1 relation between conformal metrics on $\mathbb{M}$, and classical mechanical models on the same space. Specifically the time--independent mechanical action $S$ (Hamilton's principal function) equals the conformal factor $f$, while the potential function $U$ equals minus two times the Ricci curvature of $\mathbb{M}$.}

\section{Proof of the theorem}\label{buch}

\noindent
{\it A mechanics from a given Riemannian metric.}\\
Starting from a knowledge of the metric (\ref{metrik}) on $\mathbb{M}$, we will construct a classical mechanical system having $\mathbb{M}$ as its confirguration space. We recall that, for a point particle of mass $m$ subject to a time--independent potential $U$, the Hamilton--Jacobi equation for the time--dependent action $\tilde S$ reads 
\begin{equation}
\frac{\partial \tilde S}{\partial t}+\frac{1}{2m}\left(\nabla \tilde S\right)^2+U=0.
\label{hamjacbtrev}
\end{equation}
It is well known that, separating the time variable as per 
\begin{equation}
\tilde S=S-Et,
\label{trennung}
\end{equation}
with $S$ the time--independent action (Hamilton's principal function), one obtains
\begin{equation}
\frac{1}{2m}\left(\nabla S\right)^2+U=E.
\label{stillnight}
\end{equation}
Eqn. (\ref{trennung}) suggests separating variables in (\ref{konvfluss}) as per
\begin{equation}
\tilde f=f+Et,
\label{getrennt}
\end{equation}
where the sign of the time variable is reversed\footnote{This time reversal is imposed on us by the time--flow eqn. (\ref{konvfluss}), with respect to which time is reversed in the mechanical model. This is just a rewording of (part of) section 6.4 of ref. \cite{TOPPING}, where a corresponding heat flow is run {\it backwards}\/ in time.}
 with respect to (\ref{trennung}). Substituting (\ref{getrennt}) into  (\ref{konvfluss}) leads to 
\begin{equation}
\left(\nabla f\right)^2-2\nabla^2f=E.
\label{masymas}
\end{equation}
Comparing (\ref{masymas}) with  (\ref{stillnight}) we conclude that, picking a value of the mass $m=1/2$, the following identifications can be made:
\begin{equation}
S=f, \qquad U=-2\nabla^2f = -2R.
\label{geoint}
\end{equation}

\noindent
{\it A Riemannian metric from a given mechanics.}\\
Conversely, if we are given a classical mechanics as determined by an arbitrary time--independent action $S$ on $\mathbb{M}$, and we are required to construct a conformal metric on $\mathbb{M}$, then the solution reads $f=S$. This concludes the proof.

\section{Discussion}\label{diskk}

With the 1--to--1 relation established above one can exchange a conformally flat metric for a time--independent action functional satisfying the Hamilton--Jacobi equation.
In the second part of the theorem one defines a Riemann metric, starting from the knowledge of a given mechanics. However it is not guaranteed that the metric so obtained is the {\it canonical}\/ one corresponding to the Riemann surface on which the given mechanics is defined. An example will illustrate this point.  Riemann surfaces in genus greater than 1 can be obtained as the quotient of the open unit disc $D\subset\mathbb{C}$ by the action of a Fuchsian group $\Gamma$ \cite{FARKASKRA}. As the quotient space $D/\Gamma$, the Riemann surface $\mathbb{M}$ now carries a Riemannian metric of constant negative curvature, inherited from that on the disc $D$. This hyperbolic metric is the canonical metric to consider on $\mathbb{M}$. On the other hand, the metric provided our theorem need not be hyperbolic. For example, by (\ref{geoint}) we have that the Ricci scalar curvature $R$ and the potential function $U$ carry opposite signs. Given a mechanics on $\mathbb{M}$, this determines the sign of $U$ (modulo additive constants), hence also the sign of $R$, which need not be the constant negative sign corresponding to a hyperbolic Riemann surface as explained above. However there is no contradiction. It suffices to realise that the metric induced by the mechanics considered need not (and in general will not) coincide with the hyperbolic metric induced on $D/\Gamma$ by $D$. 

Our theorem may be regarded as providing a mechanical system that is naturally associated with a given metric. Although we have  considered the {\it classical}\/ mechanics associated with a given conformal factor, one can immediately construct the corresponding {\it quantum}\/ mechanics, by means of the Schroedinger equation for the potential $U$. In fact the spectral problem for time--independent Schroedinger operators with the Ricci scalar as a potential function has been analysed in ref. \cite{TOPPING}.  We can therefore restate our result as follows: we have established a 1--to--1 relation between conformally flat metrics on configuration space, and quantum--mechanical systems on that same space. That the Ricci flow plays a key role in the quantum theory  has been shown in refs. \cite{CARROLL1, CARROLL2}. 

Moreover, the Perelman functional (\ref{epe}) also arises in the Brans--Dicke theory of gravitation, in models of conformal gravity (for a review see, {\it e.g.}, ref. \cite{CARROLL5}), and in the semiclassical quantisation of the bosonic string \cite{DHOKER}.  Further applications have been worked out in refs. \cite{NOI1, NOI2} in connection with emergent quantum mechanics \cite{ELZE1, ELZE5}. After finishing this paper we became aware of ref. \cite{KOCH}, where issues partially overlapping with ours are dealt with.

Altogether we see that the Ricci flow and the Perelman functional have important links to classical and quantum physics. Our conclusions here reaffirm the importance of these links.

\vskip.5cm
\noindent
{\bf Acknowledgements} J.M.I. is pleased to thank Max--Planck--Institut f\"ur Gravitationsphysik, Albert--Einstein--Institut (Potsdam, Germany) for hospitality extended over a long period of time.---{\it  Irrtum verl\"asst uns nie, doch ziehet ein h\"oher Bed\"urfnis immer den strebenden Geist leise zur Wahrheit hinan.} Goethe.

\end{document}